\documentclass[]{iopart}
\usepackage{graphicx}

\newcommand{\dd}{{\textrm d}}
\newcommand{\be}{\begin{equation}}
\newcommand{\ee}[1]{\label{#1} \end{equation}}
\newcommand{\ba}{\begin{eqnarray}}
\newcommand{\ea}[1]{\label{#1} \end{eqnarray}}

\begin{document}

\title[]{Pion and Kaon Spectra from Distributed Mass Quark Matter}

\author{T.~S.~Bir\'o, K.~\"Urm\"ossy and G.~G.~Barnaf\"oldi}

\address{KFKI Research Institute for Particle and Nuclear Physics,
H-1525 Budapest P.O.Box 49, Hungary }
\ead{tsbiro@sunserv.kfki.hu}

\begin{abstract}
After discussing some hints for possible masses of quasiparticles
in quark matter on the basis of lattice equation of state,
we present pion and kaon transverse spectra obtained by recombining quarks with 
distributed mass and thermal cut power-law momenta as well as fragmenting by NLO 
pQCD with intrinsic $k_T$ {and nuclear} broadening.
\end{abstract}



\section{Hadronization by quark coalescence}

Quark coalescence or recombination proves to be a remarkably simple and
powerful model of hadronization of quark matter in central, relativistic
heavy ion collisions. Transverse spectra and the constituent quark scaling of 
ellipticity in the collective flow
are predicted well by this concept, although theoretical arguments have been
raised against it time to time. In this contribution we review first a few
arguments in favor of the quark recombination model ALCOR and then present
application results to pion and kaon transverse spectra at RHIC.
We propose to consider the sliding slope of transverse spectra, with 
transverse flow correction wherever it seems to be necessary, as an excellent
indicator of deviations from exponential and as a smooth fitting edge to
high-$p_T$ spectra calculated in NLO pQCD by the use of fragmentation functions.
This sliding slope is linear in the energy in a co-flowing cell for a cut power-law
distribution, so it reflects the power-law tail nicely. The inverse powers,
the $1/(q-1)$ parameters, on the other hand should scale according to quark coalescence
ratios for mesons and baryons when produced by recombination.


A common worry with respect to recombination is represented by the question of
entropy reduction due to diminishing the color degree of freedom. Albeit the
dynamical details of any non-perturbative hadronization process are unknown to date,
there can be constraints from lattice QCD equation of state (eos) on the possible
reduction. In ref.\cite{BiroZimPLB} we have shown that lattice QCD allows for
a significant reduction in the effective number of ideal gas particles by a constant
total entropy (adiabatic process). This way the recombination process does not
necessarily reduce entropy, if at the same time an adiabatic expansion of the volume
takes place.


Another point is raised by asking why to bother statistical and thermal models of
quark matter when perturbative QCD (pQCD) is able to predict both the equation of state
and the $p_T$ spectra. The grounds here are also theoretical: on the one hand
non-perturbative methods in field theory are available for thermal systems to date,
therefore it is important to investigate whether the phenomena we observe experimentally
may be thermalized or not, on the other hand 
{pQCD might not be applicable for infrared unsafe physical 
quantities in a thermal state at all.}

\begin{figure}
\begin{center}
\includegraphics[width=0.45\textwidth,angle=-90]{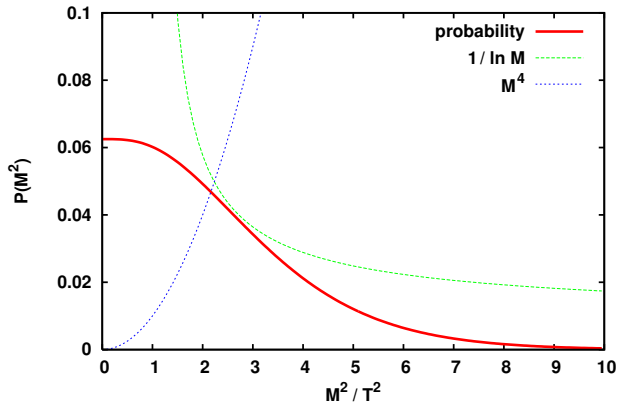}
\end{center}
\caption{\label{FIG_PM}
 The distribution of invariant mass squared for two Boltzmann distributed, massless
 partons. As examples an IR safe quantity, $M ^4$, and an unsafe one, $1/\ln(M^2)$
 are also indicated by thinner lines.
}
\end{figure}

To illustrate this,
\fref{FIG_PM} plots the probability distribution function of the invariant mass
square, $M^2$, combined from two massless, Boltzmann distributed partons:
\be
 P(M^2) = {\cal N} \int\!\dd ^3p_1 \dd ^3p_2 \, f_1 \: f_2 \: 
 \delta\left(M^2-2E_1E_2+2\vec{p}_1\cdot\vec{p}_2 \right),
\ee{PMdef}
with $p_i=(E_i,\vec{p}_i)$ four-momenta of the partons, $f_i=e^{-E_i/T}$ Boltzmann
distributions and a temperature dependent normalization factor ${\cal N}$.
The probability is normalized as $\int P(M^2) \dd M^2=1$ and due to kinematic reasons
$0 \le M^2 \le 4 E_1E_2$  for physical particles. 
The result of the integral (\ref{PMdef}) is given by
\be
 P(M^2) = \frac{1}{64} \left(\frac{M^3}{T^3} K_1(\frac{M}{T}) + 2 \frac{M^2}{T^2} K_2(\frac{M}{T}) \right).
\ee{PMresult}
This probability is finite at low $M$,
a typical infrared unsafe quantity is $1/\ln(M^2/\Lambda^2)$.


For the hadronization from a quark matter we shall consider cut power-law (Tsallis)
distributed quark momenta at a given temperature, $T$, and parameter $q$:
\be
 f_Q(\vec{p} \,) = \frac{1}{Z} \left(1 + (q-1) \frac{E(\vec{p} \, )}{T} \right)^{-\frac{1}{q-1}}.
\ee{QTsallis}
The energy - momentum dispersion relation is presumably not near to the free
one at the hadronization process. In the distributed mass model we consider
a massive dispersion relation, $E(\vec{p}\,)=\sqrt{m^2+\vec{p}\,^2}$ and a mass distribution,
$\rho(m)$. The smeared quark masses give some flexibility to recombine any hadron mass
we target at, in particular the low pion mass. We use two approaches: i) a
model mass distribution suppressing both low and high quark masses \cite{BiroJPG},
\be
 \rho(m) = a \exp\left(- \frac{m}{m_0} - \frac{m_0}{m}\right),
\ee{EXPQM}
and ii) a distribution with a mass gap deviced to mimic lattice eos data \cite{BiroPRC}
(see later),
\be
 \rho(m) = \frac{4}{\pi} \frac{m_0}{m^2} \sqrt{1-\frac{m_0^2}{m^2}}.
\ee{GAPQM}
Before presenting our results for the meson spectra, we discuss briefly what lattice eos
may hint at the existence and shape of $\rho(m)$ distributions.

\section{Quasiparticles and QCD eos}


We would like to point out that $T > T_c$ quark matter does not necessarily
behave like approaching an ideal quark-gluon plasma (QGP) at very high temperature.
The best test considered for this is the interaction measure, $e-3p$, which
vanishes for an ideal massless gas and in the $T \rightarrow \infty$
limit. Lattice gauge theory and QCD data all agree with this expectation.

However, a different question is the high-temperature scaling of this quantity.

\begin{figure}
\begin{center}
\includegraphics[width=0.45\textwidth,angle=-90]{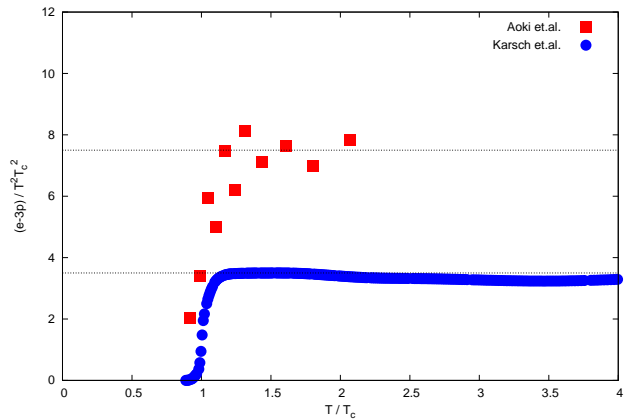}
\end{center}
\caption{\label{FIG_IM}
 Interaction measure scaled by $T^2$ for pure SU(3) lattice gauge theory \cite{Karsch} and
 for 2+1 flavor QCD with physical pion mass obtained recently \cite{Aoki}.
 The constant high-T asymptotic of this quantity may correspond to a constant mass
 \cite{BiroVanLaszlo} 
 or to screened strings contributing to the free energy density as $n^{2/3}$ 
 \cite{BiroShanTon}.
}
\end{figure}

\Fref{FIG_IM} plots the interaction measure scaled by the square of the 
temperature ({not, as usual, by $T^4$}) for pure SU(3) lattice eos
\cite{Karsch} and for a more recent 2+1 flavor QCD eos with physical pion mass \cite{Aoki}.
Both seem to agree with a constant for 
\be
 \frac{e-3p}{T^2} = T^3 \frac{\dd}{\dd T} \left(\frac{p}{T^4} \right) 
\ee{e3p}
soon after the color deconfinement
up to the highest temperature investigated on the lattice.

Not any model of quark matter satisfies this constraint. 
In the quasiparticle picture for an ideal gas of mass $m$ particles
the $T\rightarrow \infty$ limit is just $m^2/2$, in case of a mass
distribution it is given by $\langle m^2 \rangle/2 - \dd / \dd T(\langle m^2\rangle)/(4T^5)$
to leading order.
While this may well be a constant for some temperature dependent mass distribution $\rho(m;T)$, the 
pQCD leading term, $T^3 \dd / \dd T(1-a_2g^2) = (a_2/\alpha_0)\:  g^2T^2$ for the usual one loop
expression $g^2(T)=\alpha_0/\ln(T^2/\Lambda^2)$, 
is not a constant\footnote{although also proportional to the
thermal mass squared.}.



In principle in advanced quasiparticle models the mass smearing may also be
temperature dependent. In this case a mean field term occurs in the pressure,
which nevertheless cancels in the combination $e+p$. Integrals of the PDF
$\rho(m)$ leading to a given $e+p$ can be compared to lattice eos, too.
In this case the Markov inequality gives estimates for the lowest mass\cite{BiroVanLaszlo}.

\section{Hadron $p_T$ spectra from quark matter}

\begin{figure}
\begin{center}
\includegraphics[width=0.45\textwidth,angle=-90]{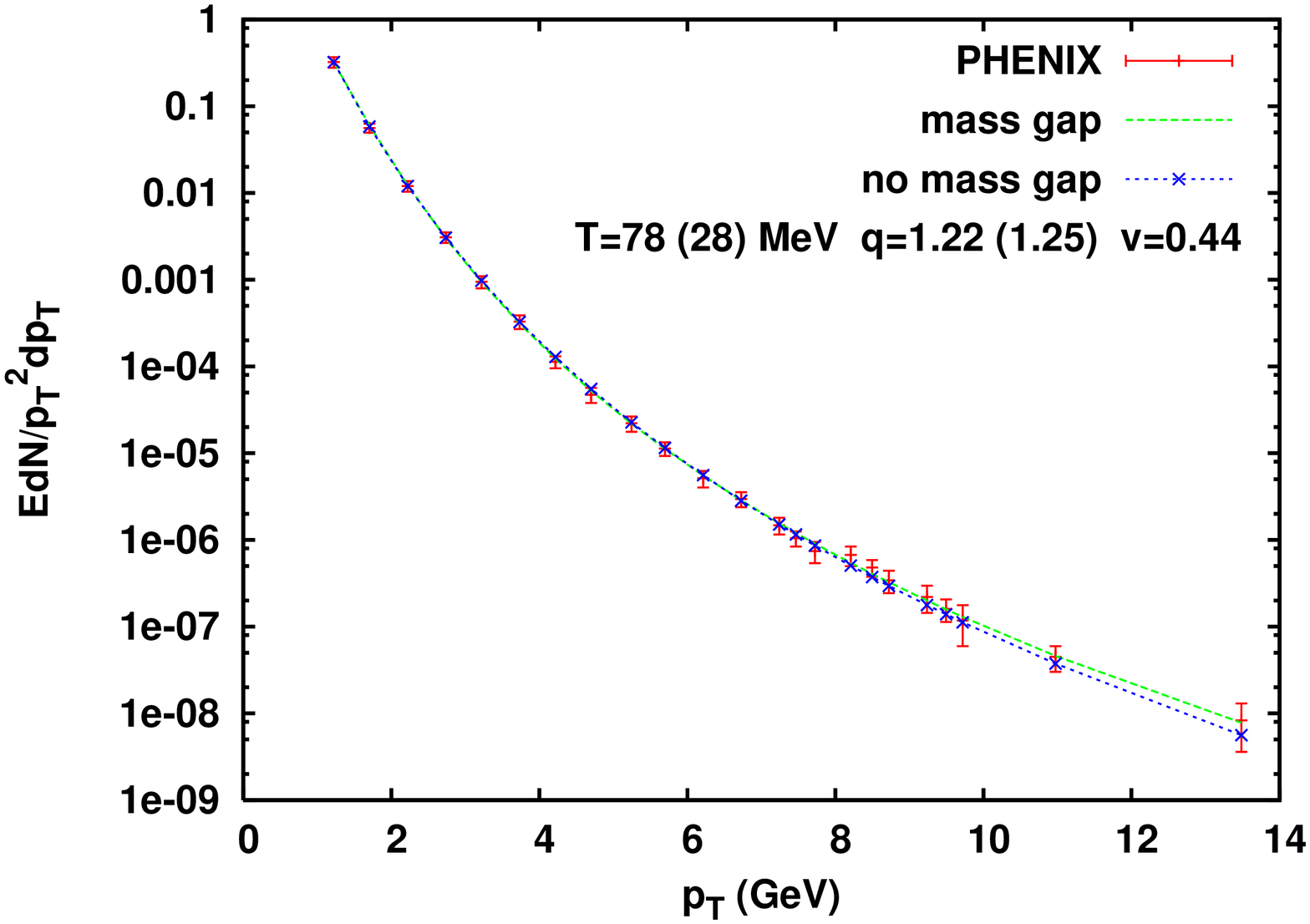}
\includegraphics[width=0.45\textwidth,angle=-90]{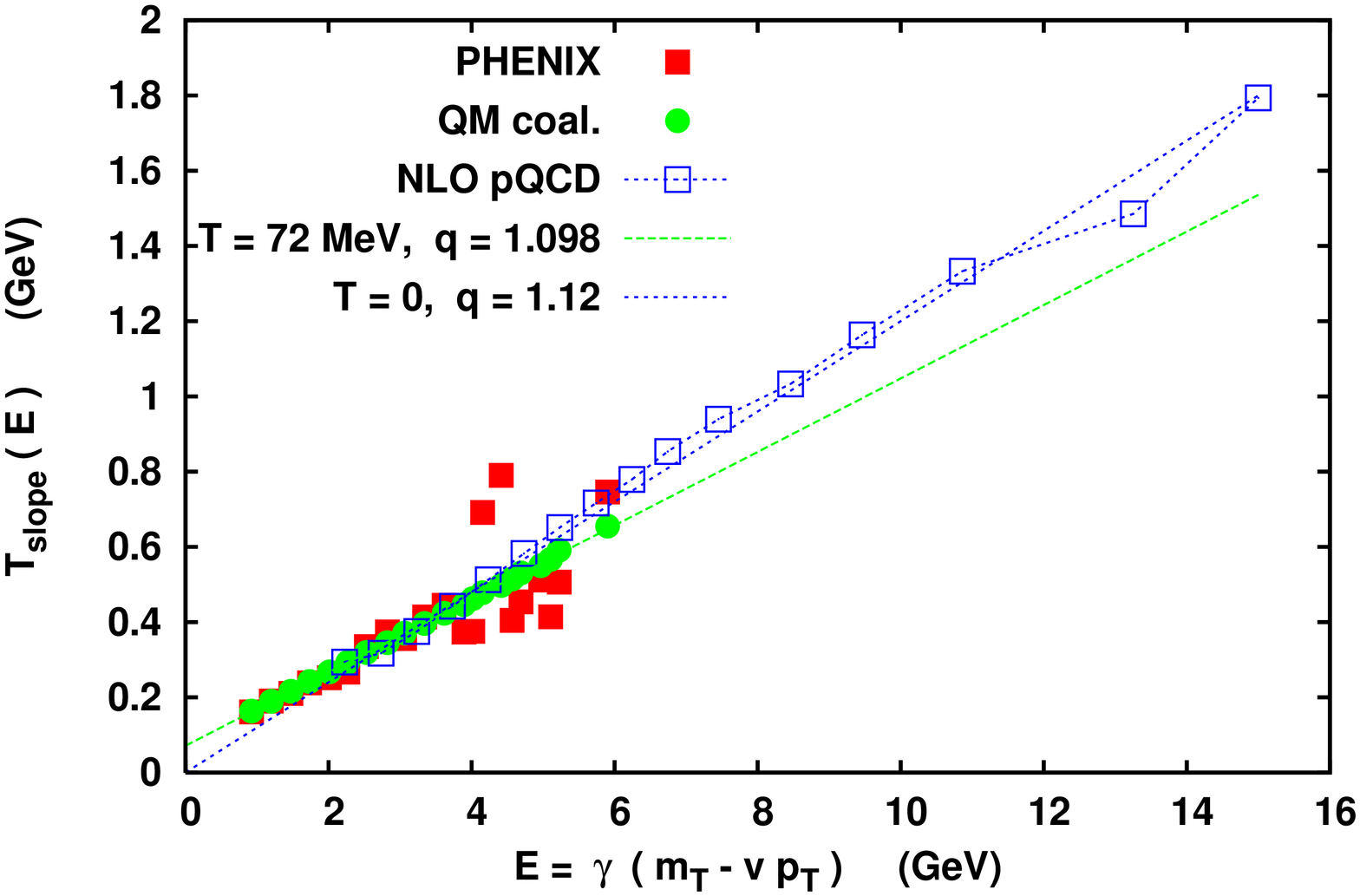}
\end{center}
\caption{\label{FIG_PION}
 Energy distribution in a blast wave's rest frame for pion $p_T$ spectra (upper) and
 the sliding slope obtained from it (lower). 
}
\end{figure}

\begin{figure}
\begin{center}
\includegraphics[width=0.45\textwidth,angle=-90]{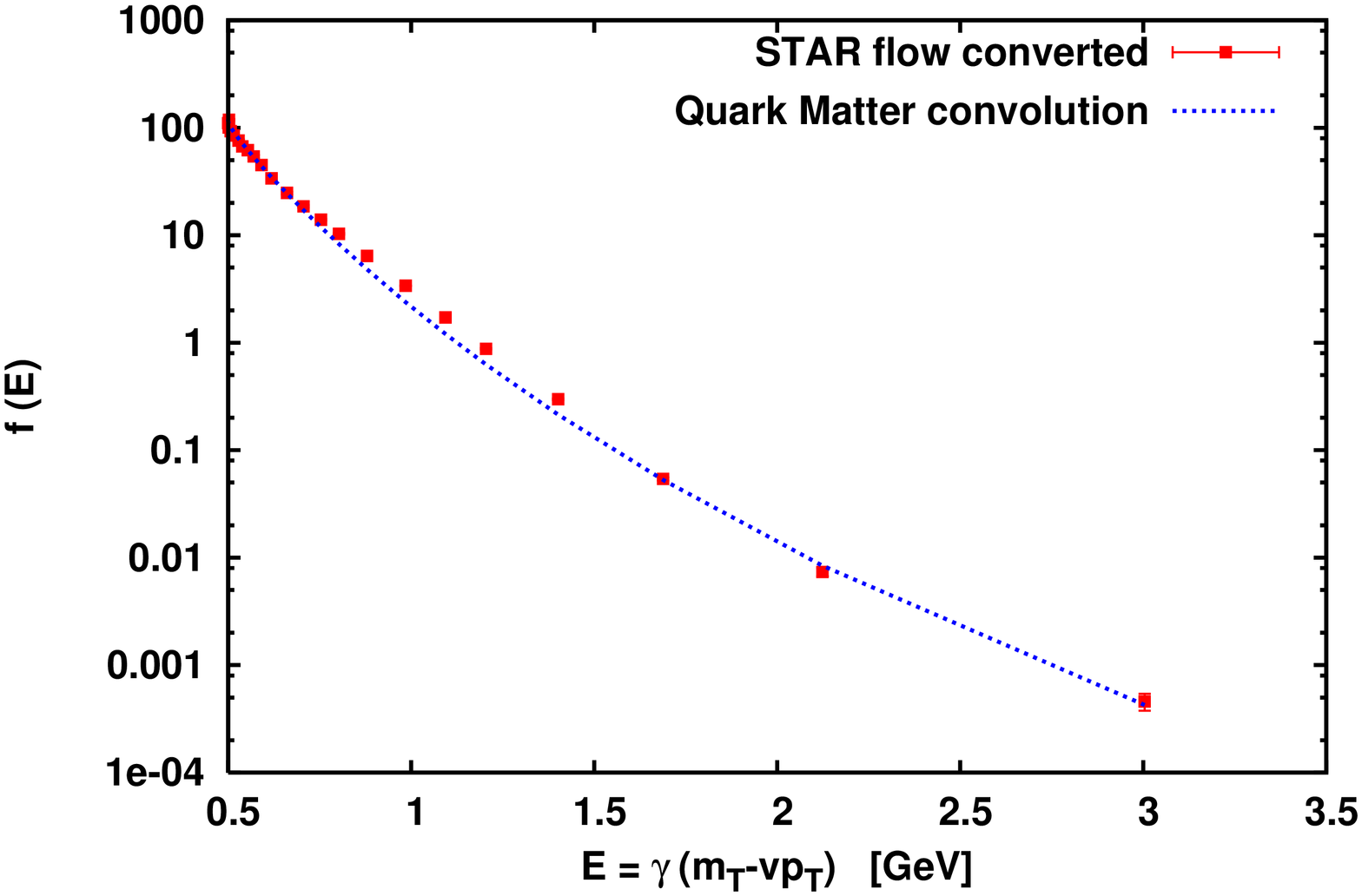}
\includegraphics[width=0.45\textwidth,angle=-90]{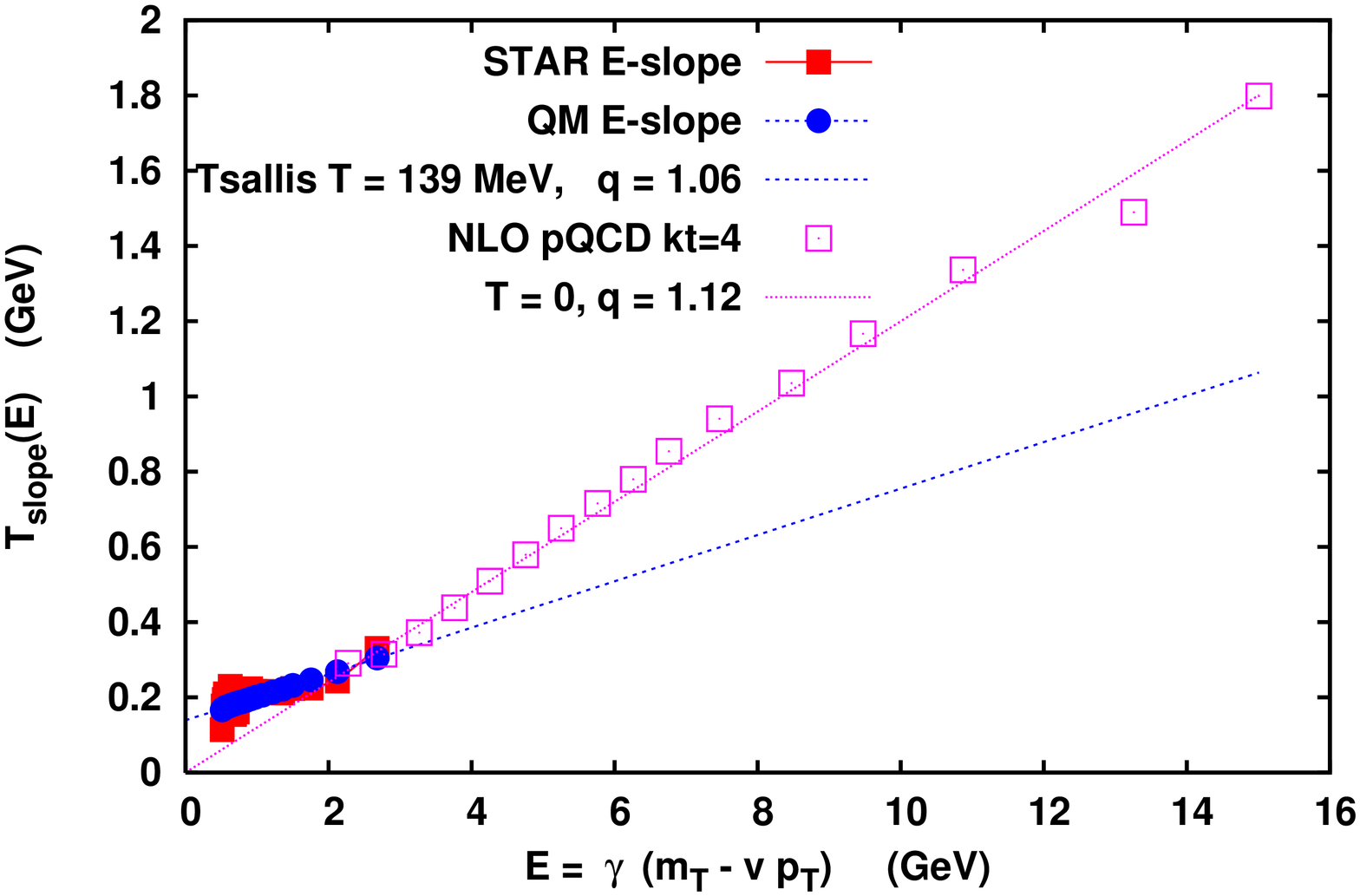}
\end{center}
\caption{\label{FIG_KAON}
 Conjectured co-moving energy distribution from STAR\cite{STAR} kaon $p_T$ data 
(upper) and the sliding slope calculated from it (lower). Theoretical points from 
quark matter recombination and lines from NLO pQCD fragmentation calculation are 
plotted. The lower line is without intrinsic $k_T$ assumption, for the upper line 
we assumed $\langle k_T^2 \rangle = 4$ GeV$^2$/c$^2$.
}
\end{figure}

Finally we present transverse momentum spectra for pions and kaons as obtained from
quark recombination and pQCD fragmentation. \Fref{FIG_PION} and \fref{FIG_KAON}
show for pions and kaons the respective curves. The upper patterns are invariant
yields, $f(E)$, while the lower pictures are the sliding slopes
\be
 T_{{\rm slope}} \: = \: -  \, \left(\frac{ \dd }{\dd E} \ln f(E) \right)^{-1}.
\ee{SLOPE}
Here $E = \gamma(m_T-vp_T)$ is the average transverse flow corrected energy
at zero rapidity.
We took into account a transverse flow of $v\approx 0.5$ for the kaons.
For a Tsallis distribution $T_{\rm slope}(E) = T^* + (q-1) E$, the steepness of the
linear rise is not affected by the blue shift factor, 
but the intersect is: $T^*=T\sqrt{(1+v)/(1-v)}$ for $E \gg m$.

{ We used a next-to-leading-order pQCD-improved parton model for 
calculating the high-$p_T$ pion and kaon spectra in $AuAu$ collisions. 
Theoretical calculations presented here, are based on 'standard' parton distribution 
functions~\cite{PDF} (PDF), but in a generalized way: a $2$-dimensional Gaussian transverse 
momentum distribution were included, represented by the average intrinsic transverse 
momenta, $\langle k_T^2 \rangle $ (see detail in Refs.~\cite{pQCD1,pQCD2}). The 
calculation were carried out both without (standard pQCD calculation) and with 
intrinsic-$k_T$, as motivated by experimental data~\cite{Zielinsky,PhenixKT}.}      

{The calculated spectra strongly depend on initial and final state nuclear 
effects, which have to be taken into account in proton-nucleus and 
nucleus-nucleus  collisions. Here we used the nuclear shadowing 
and multiple scattering, and the jet-quenching with the corresponding opacity 
parameter value, $L/\lambda \approx 3.5 - 4.0$.}  

{We applied two parameterizations for fragmentation: KKP~\cite{KKP} and 
the latest version of AKK~\cite{AKK}. 
}

In conclusion the curves from thermal cut power-law distributed  quark matter 
recombination meet smoothly those NLO pQCD results, which take into account an 
intrinsic transverse momentum in the parton distributions. The $(q-1)$ steepness 
of the energy dependence of the inverse logarithmic slope, $T_{{\rm slope}}$ is 
expected to follow a coalescence rule for mesonic and baryonic recombination at 
high $p_T$ only.

\ack{ This work has been supported by the Hungarian National Fund OTKA
(T49466, NK62044 and IN71374). One of the authors (GGB) would 
like to thank also for P. L\'evai, G. F\'ai and G. Papp for discussions and 
the pQCD code.}


\end{document}